\documentclass[10pt,conference]{IEEEtran}
\IEEEoverridecommandlockouts
\usepackage{cite}
\usepackage{amsmath,amssymb,amsfonts}
\usepackage{algorithm}
\usepackage{algpseudocode}
\usepackage{graphicx}
\usepackage{subfigure}
\usepackage{textcomp}
\usepackage{xcolor}
\usepackage{stfloats}
\def\BibTeX{{\rm B\kern-.05em{\sc i\kern-.025em b}\kern-.08em
    T\kern-.1667em\lower.7ex\hbox{E}\kern-.125emX}}
\graphicspath{{Figures/}{./}}
\usepackage[numbers,sort&compress]{natbib}

\renewcommand{\baselinestretch}{1}
\begin{document}

\algnewcommand\algorithmicswitch{\textbf{switch}}
\algnewcommand\algorithmiccase{\textbf{case}}
\algnewcommand\algorithmicassert{\texttt{assert}}
\algnewcommand\Assert[1]{\State \algorithmicassert(#1)}%
\algdef{SE}[SWITCH]{Switch}{EndSwitch}[1]{\algorithmicswitch\ #1\ \algorithmicdo}{\algorithmicend\ \algorithmicswitch}%
\algdef{SE}[CASE]{Case}{EndCase}[1]{\algorithmiccase\ #1}{\algorithmicend\ \algorithmiccase}%
\algtext*{EndCase}%

\title{Performance of Cell-Free Massive MIMO in Realistic Urban Propagation Environments

}


\author{\IEEEauthorblockN{Yunlu~Xiao
and
Ljiljana~Simić}
\IEEEauthorblockA{\textit{Institute for Networked Systems, 
RWTH Aachen University},
Aachen, Germany\\ Email: \{yxi, lsi\}@inets.rwth-aachen.de}}

\maketitle

\begin{abstract}
While UE-centric cell-free massive MIMO (\mbox{CF-mMIMO}) provides high and uniform throughput performance under the assumption of a uniform propagation environment modeled by the log-distance path loss channel model, the performance under a realistic urban propagation environment is not yet fully addressed. In this paper we conduct the first comparative performance study of CF-mMIMO under both the widely assumed log-distance channel model and the realistic urban propagation environment obtained via raytracing using real 3D city layouts and practical AP locations. Our results show that with the raytracing channel model, CF-mMIMO cannot achieve as high and uniform throughput performance as observed with the log-distance channel model, putting into question the attractiveness in practice of CF-mMIMO for real urban depolyments.
\end{abstract}

\vspace{-0.1cm}
\section{Introduction}
Cell-free massive multiple-input-multiple-output (\mbox{CF-mMIMO}) has been proposed to achieve both high and uniform throughput performance to the whole network \cite{cfvs} by coordinating all access points (APs) in the network to jointly serve all UEs. In this case, no UE will become a ``cell-edge'' UE due to low desired signal strength from the serving AP and high interference from other APs, since every AP serves the UE. However, managing the entire network to simultaneously serve a given UE brings unlimited signaling overhead and system cost \cite{scalableCF}. UE-centric \mbox{CF-mMIMO} is thus proposed \cite{cfsim} to limit the serving set size while still reducing the edge-effect. The network is divided into disjoint clusters managed by central processing units (CPUs) and the APs are assigned to the CPU clusters based on their locations. To form the limited serving set, the UE first selects a few APs with the best channels, and then the APs in the clusters that those APs belong to cooperate to serve this UE. Under the widely adopted assumption of randomly uniformly distributed APs according to a Poisson point process (PPP) and a log-distance path loss channel model, the UE-centric CF-mMIMO architecture achieves good performance, especially for the worst-performing UEs. This is because, under these two idealized assumptions, each UE is effectively surrounded by its serving AP set, which effectively provides uniformly good channel conditions for all UEs \cite{scalableCF, cfsim, cfbook, cfl3, cfParadigm}.

However, the layout of realistic network deployments and the corresponding propagation environment in an urban area are site-specific and cannot be accurately modeled by PPP distributions and the log-distance channel model. Firstly, the location of APs and UEs are constrained by buildings, streets, and public squares. Furthermore, according to the \mbox{log-distance} path loss model, the wireless signal strength strictly depends on the distance between the AP and UE. This is not the case in real urban wireless channels, where the propagation mechanisms of scattering, reflection, and diffraction give rise to building shadowing and ``urban canyon'' effects, resulting in site-specific, non-uniform path loss variation with AP-UE distance. Consequently, in contrast to the simplistic log-distance path loss model, in realistic urban propagation environments we may observe a significant difference in the channel quality from two APs at a similar location or distance to the UE. Therefore, in the realistic urban propagation environment, the AP serving set of UE-centric CF-mMIMO may no longer be able to provide a uniformly good channel to all UEs, such that the ``cell-edge'' effect of relatively poorly covered UEs may persist in practice. It is thus uncertain whether the key benefit of high and uniform throughput performance from the classical UE-centric CF-mMIMO literature can in fact still be achieved in a realistic urban network.

To the best of our knowledge, the majority of the literature on CF-mMIMO performance analysis \cite{cfvs, scalableCF, cfsim, cfbook, cfl3, cfParadigm} only considers a random uniform AP/UE distribution and the log-distance path loss model. The works in \cite{poster, beerten2023cellfree} take into account the \mbox{site-specific} AP locations in cities, but the path loss is still modeled by a log-distance model. The works in \cite{Cfmeasure, cfRealmodel, 2022Evaluation} study the site-specific channel corresponding to CF-mMIMO systems in urban environments via measurement or raytracing, but do not consider any network-level throughput performance analysis. Thus, a fundamental question must be revisited for the realistic urban propagation environment: is UE-centric CF-mMIMO an attractive network architecture in \textit{practical} urban networks? \looseness=-1
 
To address this question, we conduct the first comparative performance analysis of CF-mMIMO under both the widely assumed log-distance channel model and the realistic \mbox{site-specific} urban propagation environment obtained via raytracing using real 3D city layouts and practical AP locations. We present a comprehensive study of the performance of \mbox{CF-mMIMO} under the two channel models for two representative urban environments in Amsterdam and Dresden, considering different network architectures in terms of serving AP selection and CPU cluster divisions. Our results show that with the raytracing channel model representing the realistic urban propagation environment, \mbox{CF-mMIMO} cannot achieve the high throughput performance as observed under the \mbox{log-distance} channel model assumption, especially for the \mbox{worst-performing} UEs. This puts into question whether \mbox{CF-mMIMO} would in practice deliver the promise of high and uniform performance in real urban networks.


\section{System Model}
\label{model}

\subsection{Network Topology}
\label{locate}
\begin{figure}[!tb] 
\vspace{0.1cm}
	\centering
	\subfigure[city of Amsterdam]{
		\label{city.sub.1}
		\includegraphics[width=0.48\linewidth]{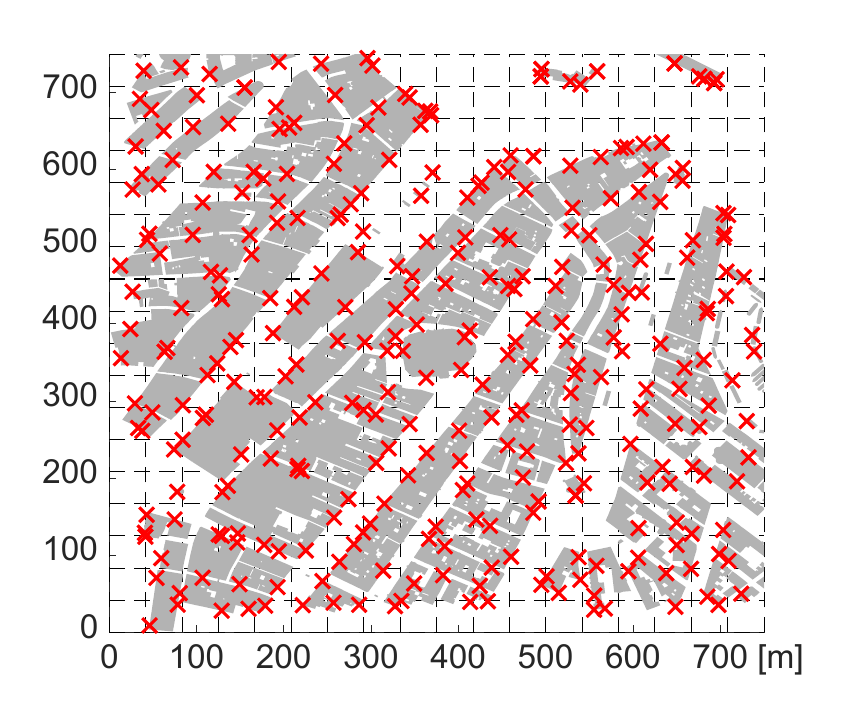}}
	\subfigure[city of Dresden]{
		\label{city.sub.2}
		\includegraphics[width=0.48\linewidth]{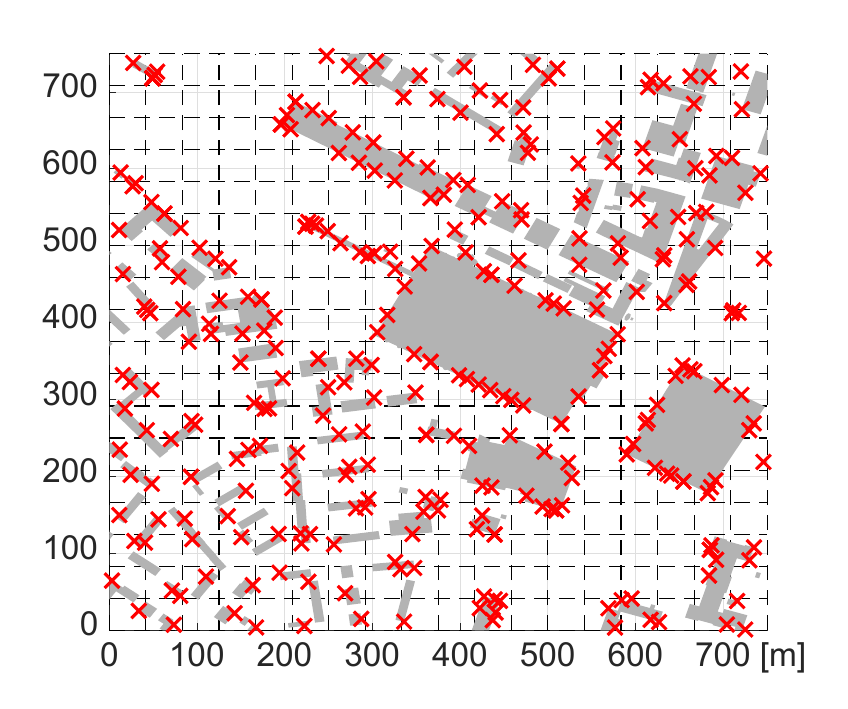}}
	\caption{Urban AP distributions according to Algorithm \ref{aplocate} for $M=324$ APs, with APs marked by crosses and the uniform grid in Step 4 of Algorithm \ref{aplocate} marked by dashed lines.}
	\label{city}
\end{figure}
\begin{algorithm}[t]
\caption{AP Placement Algorithm} \label{aplocate}
\textbf{Input:} publicly available 3D building layouts, the city area $S$ the desired AP number $M$ ($M$ as a square number).
\begin{algorithmic}[1]

\State Find the building edge and corner coordinates from the building layouts, insert coordinates every $d_1$ meter as candidate AP locations $\mathcal{L}_1$.

\State Delete the candidate locations that are too close. If 2 coordinates in  $\mathcal{L}_1$ have distance smaller than $d_2$, delete the second one. The remaining coordinates become the candidate AP location $\mathcal{L}_2$

\State Delete the coordinates that inside building court yards and small alleys in $\mathcal{L}_2$, obtain the final candidate AP locations $\mathcal{L}$.

\State General a $\sqrt{M} \times \sqrt{M}$ uniform grid in the area $S$, obtain the coordinate of the center of each grid as a set $\mathcal{C}$. The coordinate of the center of grid $m$ denoted as $C[m]$.

\For {$m = 1,...,M$}

\State Find the closest coordinate $L_{min}$ in $\mathcal{L}$ to $C[m]$.

\State Set the $m$ AP location $A_M[m] = L_{min}$.

\EndFor

\end{algorithmic}
\textbf{Output:} The AP locations set $A_M$ of the AP number $M$.
\end{algorithm}
We consider a static network with $M$ single antenna APs with height $h_{\text{AP}}=11$ m and $K$ single antenna UEs with height $h_{\text{UE}}=1$ m distributed in a city area $S$. We consider two real city areas in Amsterdam and Dresden (\textit{cf.} Fig. \ref{city}) with qualitatively dissimilar layouts, representing different urban build-up densities. We propose an AP placement algorithm in Algorithm \ref{aplocate} to determine the feasible AP locations based on the two considered realistic city layouts, distributing the APs as uniformly as possible to best achieve the ideal ``cell-free'' deployment \cite{cfvs}. We ensure the AP location to be practical by only placing the APs at edges or corners of the building. Algorithm \ref{aplocate} requires the building layouts as an input, which can be obtained via a public open data base as shapefiles \cite{shp}. We first obtain the candidate AP locations in Steps 1 to 3. In Step 1, all candidate AP locations are distributed at the edge or corner of the building blocks. In Step 2, to avoid interference and waste of infrastructure resources, the APs should not be too close to each other. Furthermore in Step 3, the APs will not be located in the inner building courtyards and small alleys, where the UEs are not likely to appear. In Steps 4 to 8, a subset of $M$ APs are sampled from the candidate locations. Fig. \ref{city} shows the resulting AP locations of the considered city layouts, with $d_1=10$ m, $d_2 = 5$ m, and $M=324$ ($\lambda = 576 \text{AP/km}^2$). For the UE locations, we first obtain the candidate UE locations sampling by a 10 m $\times$ 10 m grid in the area $S$ (consistent with our raytracing simulation granularity, \textit{cf.} Sec. \ref{channel}), excluding the locations inside buildings. Then we randomly choose $K=50$ UEs in $S$ for each Monte Carlo run for the performance analysis in Sec. \ref{results}. 


\subsection{CF-mMIMO Network Architecture}
\label{Arch}
\begin{figure}[!tb]
\vspace{0.1cm} 
	\centering
	\includegraphics[width=0.85\columnwidth]{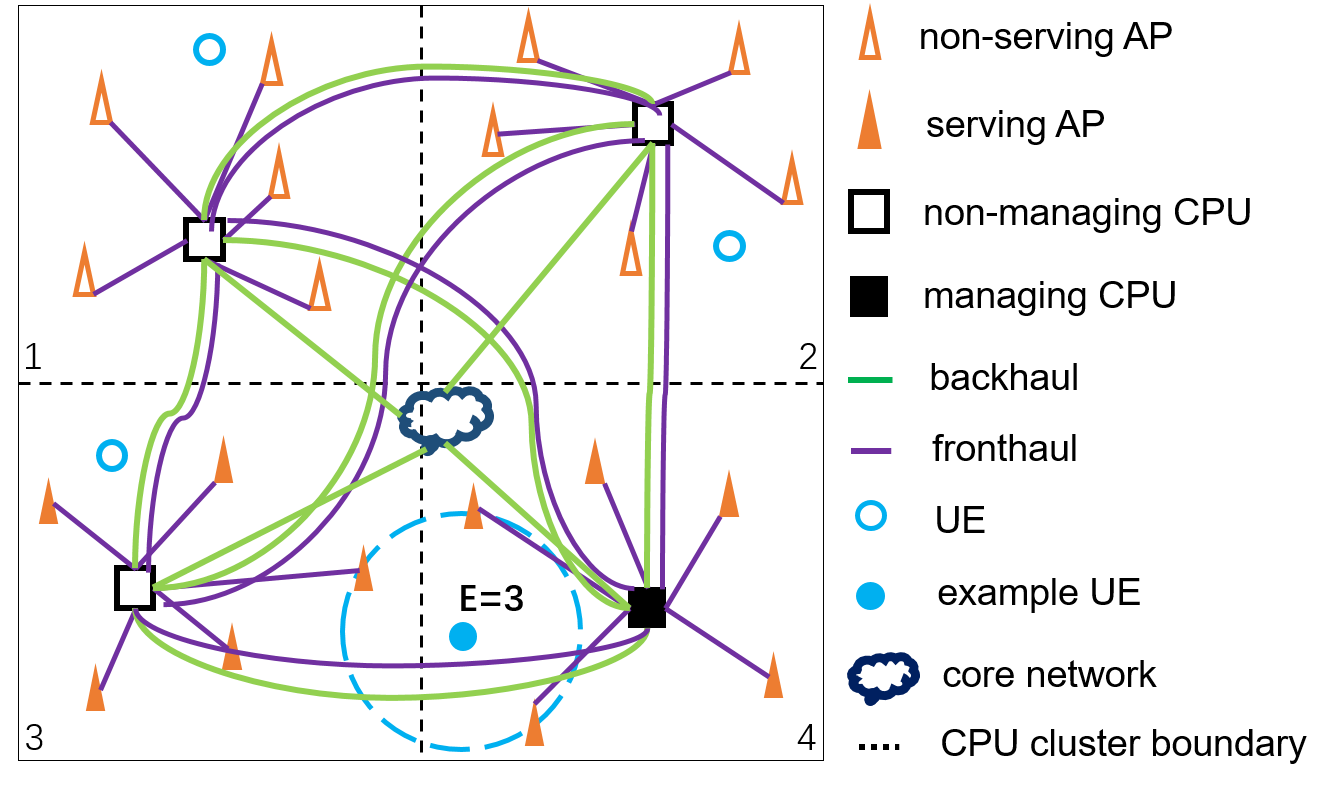}
	\caption{UE-centric CF-mMIMO network architecture.}
	\label{arch}
\end{figure}
We study the architecture of UE-centric CF-mMIMO with multiple CPUs \cite{cfsim}. The APs are assigned to disjoint CPU clusters, each with one CPU and on average $Q$ APs. Fig. \ref{arch} shows the network architecture with $Q=5$. The APs in one CPU cluster connect to the CPU via fronthaul links, while all CPUs connect to the core network via backhaul. The CPU performs the centralized signal processing for the APs and UEs residing in its cluster. Fig. \ref{arch} furthermore shows the serving AP set selection for an example UE. The UE first selects $E=3$ APs that have the best channel conditions (dashed circle), and the CPU clusters that those APs belong to (cluster 3 and 4) cooperate to serve this UE \cite{myPIMRC}. When $E=1$, the given UE is served by only one CPU cluster, which is in fact the architecture of traditional network-centric distributed MIMO \cite{comp}.  We will study different architectures in terms of $\{Q, E\}$ combinations in Sec. \ref{results}.



\subsection{Signal Processing \& Channel Model}
\label{channel}

We consider the downlink, with a coherence block with a total length of $\tau_c$ where the channel measurement and signal processing only happen every $\tau_c$ slots. We assume a randomly assigned pilot sequence occupying $\tau_p$ slots in each coherence block and sufficient pilots for each UE to avoid pilot contamination. We assume centralized signal processing CF-mMIMO transmission and minimum \mbox{mean-square} error (MMSE) decoding as in \cite{scalableCF}. We assume equal power allocation in the downlink, where each AP divides its power equally among the served UEs. We use the CF-mMIMO throughput model in \cite{scalableCF} and obtain the downlink spectral efficiency (SE) obtained with (34) in \cite{scalableCF}. We use the average signal-to-noise ratio (SNR) $\beta_{mk}$ between AP $m$ and UE $k$ to represent the channel quality, which is 
\begin{equation}
\beta_{mk} = p_{mk} / (\mathrm{L}_{m k} n_{0})
\label{eqsnr}
\end{equation}
\noindent where $p_{mk}$ is the transmit power from AP $m$ to UE $k$, $\mathrm{L}_{m k}$ is the path loss, and $n_0$ is the Gaussian noise. 
In this paper, we consider two channel models to determine the path loss $L_{mk}$: (i) raytracing, as a deterministic channel model realistically representing the site-specific wireless propagation environment of the urban network deployment area; and (ii) log-distance path loss model, as a more simplistic statistical channel model. We obtain the raytracing channel data using the Wireless Insite software from RemCom \cite{insite}, based on the real city layout (\textit{cf.} Fig. \ref{city}) and the simulation settings summarised in Table \ref{para}. As a reference, we consider the widely adopted \mbox{three-slope} \mbox{log-distance} model \cite{cfvs}, which gives the path loss with shadowing as $\mathrm{L}_{m k} = \overline{\mathrm{L}}_{m k} 10^{\sigma z_{mk} /10}$, with $\sigma$ as the shadowing deviation, $z_{mk} \sim N(0,1)$, and the average path loss in dB as:
\begin{equation}
\overline{\mathrm{L}}_{m k}=\left\{\begin{array}{l}
L_0+35 \log _{10}\left(d_{m k}\right), d_{m k}>d_1 \\
L_0+15 \log _{10}\left(d_c\right)+20 \log _{10}\left(d_0\right), d_{m k} \leq d_0 , \\
L_0+15 \log _{10}\left(d_c\right)+20 \log _{10}\left(d_{m k}\right), \text{else} \\
\end{array}\right.
\label{log}
\end{equation}
\noindent where
\begin{equation}
\begin{aligned}
L_0 = & 46.3+33.9 \log _{10}(f)-13.82 \log _{10}\left(h_{\mathrm{AP}}\right) \\
& -\left(1.1 \log _{10}(f)-0.7\right) h_{\mathrm{UE}}+\left(1.56 \log _{10}(f)-0.8\right),
\end{aligned}
\end{equation}
\noindent where $f$ is the carrier frequency, $d_c=50$ m, and $d_0=10$ m.
\begin{table}[!tb]
\caption{Simulation Parameters}
\centering
\begin{tabular}{ll}
\hline
\textit{parameter}              & \textit{value}                \\ \hline
noise figure           & 9 dB                 \\
shadowing deviation, $\sigma$    & 8 dB \\
AP transmit power      & 20 dBm               \\
carrier frequency, $f$      & 2 GHz                 \\ \hline
\textit{wireless insite parameter}   & \textit{value}           \\ \hline
antenna                & Half-wave dipole     \\
building material      & ITU Concrete 2.4 GHz \\
propagation model      & X3D raytracing       \\
number of reflections  & 6                    \\
number of diffractions & 1                    \\
outage threshold   & -250 dBm            \\ \hline
\end{tabular}
\label{para}
\end{table}

\begin{figure}[!tb] 
	\centering
	\includegraphics[width=1\columnwidth]{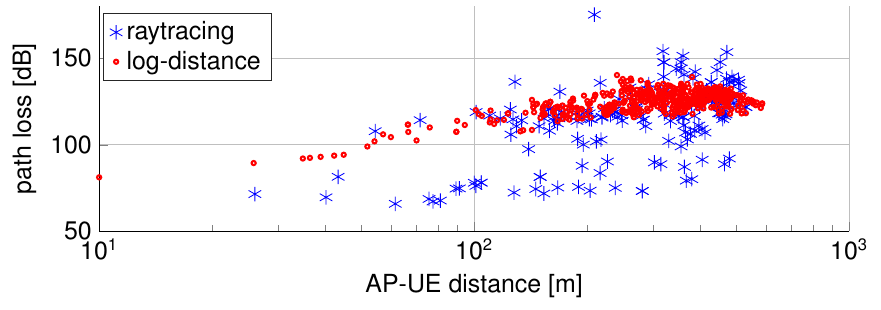}
	\caption{Path loss vs. distance for the example AP in the city of Amsterdam in Fig. \ref{snr}, with different path loss models.}
	\label{PLscatter}
\end{figure}

\begin{figure}[!tb] 
	\centering
	\subfigure[Amsterdam, raytracing channel]{
		\label{snr.sub.2}
		\includegraphics[width=0.48\linewidth]{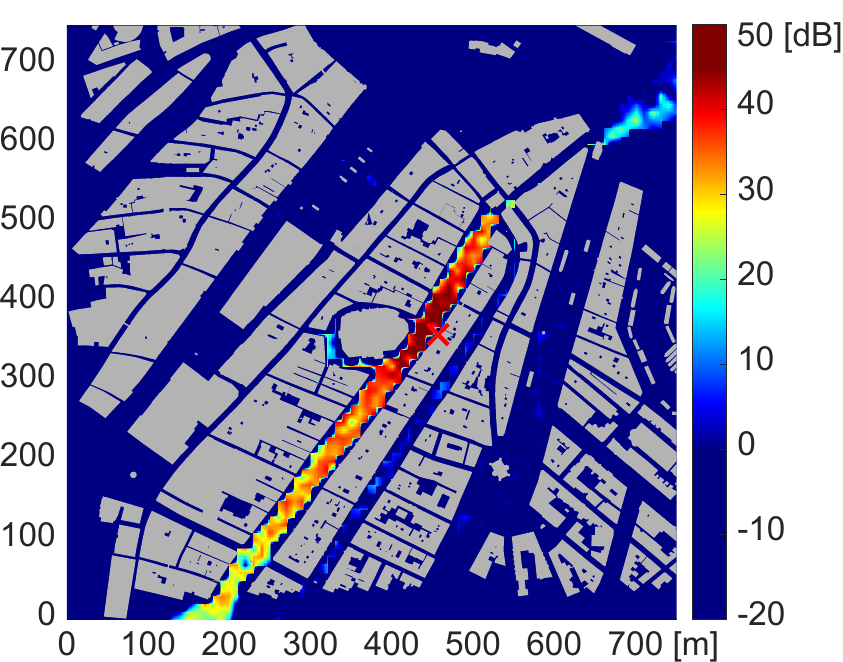}}
	\subfigure[Amsterdam, log-distance channel]{
		\label{snr.sub.1}
		\includegraphics[width=0.48\linewidth]{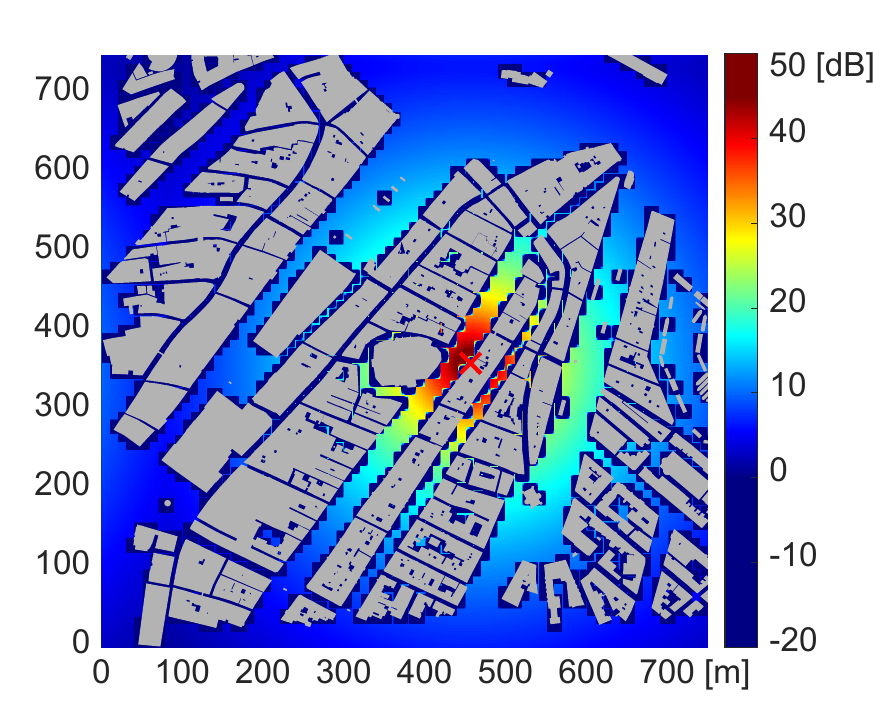}}
	\subfigure[Dresden, raytracing channel]{
		\label{snr.sub.4}
		\includegraphics[width=0.48\linewidth]{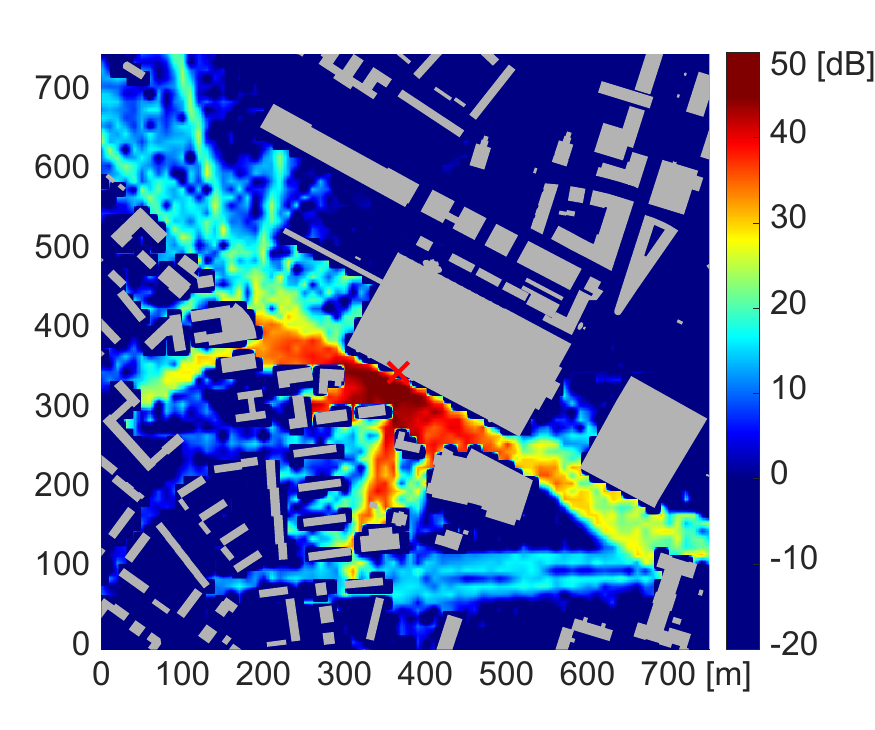}}
    \subfigure[Dresden, log-distance channel]{
		\label{snr.sub.3}
		\includegraphics[width=0.48\linewidth]{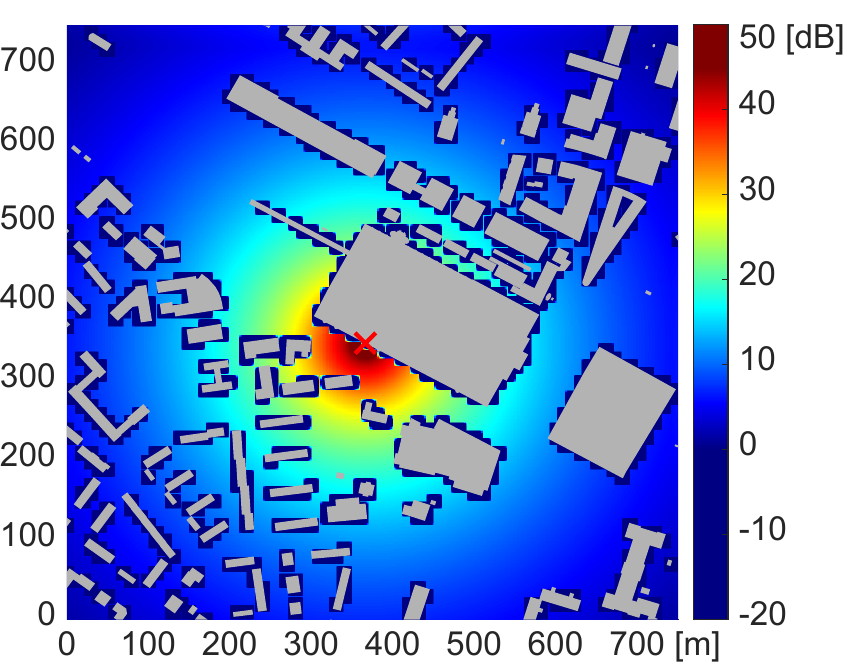}}
	\caption{Heatmap of SNR spatial distribution for different cities with different channel models, for an example AP (red cross).}
	\label{snr}
\end{figure}

Fig. \ref{PLscatter} shows the path loss versus distance for an example AP in Amsterdam for the two channel models. For the log-distance model, the path loss between AP $m$ and UE $k$ increases with the increase of distance $d_{mk}$ according to (\ref{log}). The average path loss in (\ref{log}) effectively fits the path loss of the raytracing model. However, many path loss values in the raytracing model are significantly different, with a variation of 20-40 dB from the \mbox{log-distance} model for a comparable distance, illustrating that the site-specific path loss with raytracing is far from being strictly distance-dependent. To study the propagation environment in detail, Fig. \ref{snr} shows the spatial distribution of SNR of an example AP to all candidate UE locations (10 m $\times$ 10 m grid) in Amsterdam and Dresden with different channel models. Figs. \ref{snr.sub.1} and \ref{snr.sub.3} illustrate that with the log-distance model, the SNR decreases with the increase of distance towards the example AP consistent with (\ref{eqsnr}). By contrast, Figs. \ref{snr.sub.2} and \ref{snr.sub.4} show that the SNR of the raytracing model is highly non-uniform and heavily impacted by the building distributions. For the city of Amsterdam that has large building blocks and narrow streets, Fig. \ref{snr.sub.2} shows that the AP can effectively transmit the signal even to \mbox{far-away} UEs on the same street as the AP (i.e., the ``urban canyon'' effect), whereas at other locations, there is almost no signal received due to severe building shadowing. The area in Dresden has a contrasting building layout to Amsterdam, i.e., with small building blocks and large open areas. In this type of urban environment, Fig. \ref{snr.sub.4} shows that in the open area very close to the example AP, the SNR distribution is similar to that of the log-distance model. Nonetheless, the SNR distribution over the entire network is also highly \mbox{non-uniform} and impacted by buildings. For example, at the area around $[x,y] = [100, 300]$, the received SNR from the AP is around 25 dB due to a diffraction path. However, a UE at a location very close to this path can only receive SNR below -20 dB (below a typical decoding threshold) due to building shadowing. Fig. \ref{snr} thus illustrates that the two channel models lead to a significantly different spatial distribution of SNR, with an expected knock-on effect on the throughput performance of CF-mMIMO, which we will study in detail in Sec. \ref{results}.

\section{Results}
\label{results}
\begin{figure*}[!tb] 
	\centering
	\includegraphics[width=1.1\columnwidth]{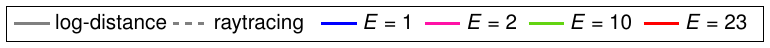}
	\subfigure[Amsterdam]{
		\label{seQ5.sub.1}
		\includegraphics[width=0.48\linewidth]{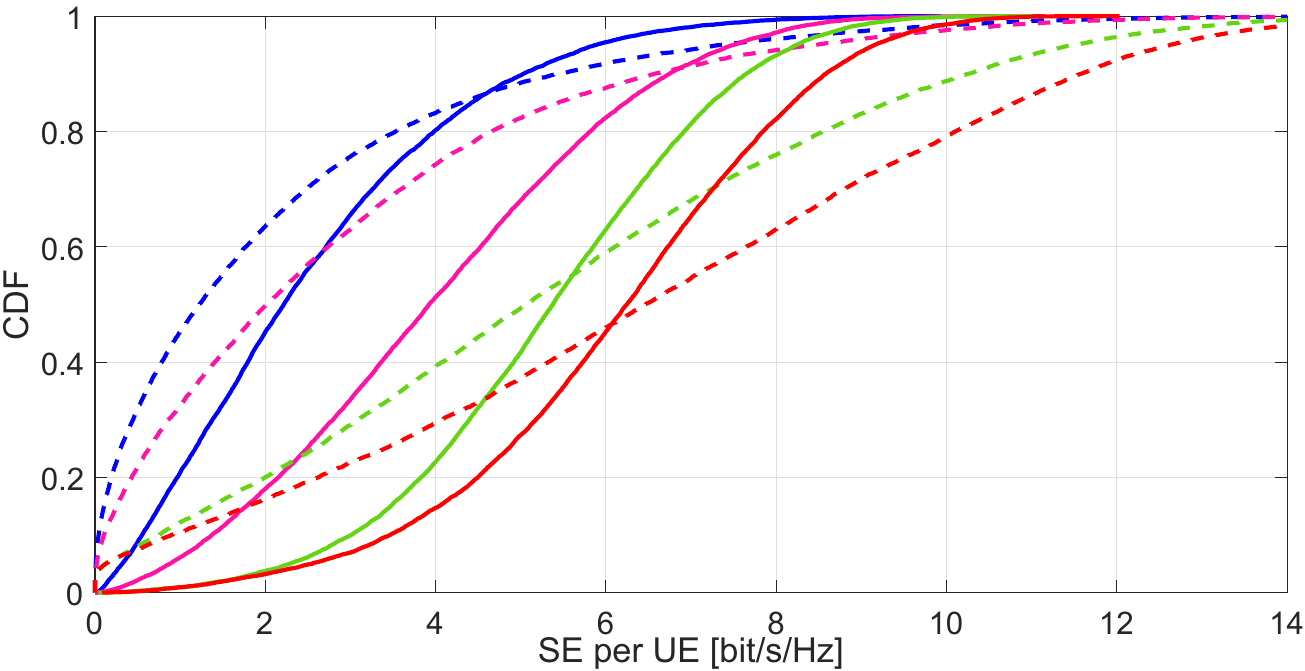}}
	\subfigure[Dresden]{
		\label{seQ5.sub.2}
		\includegraphics[width=0.48\linewidth]{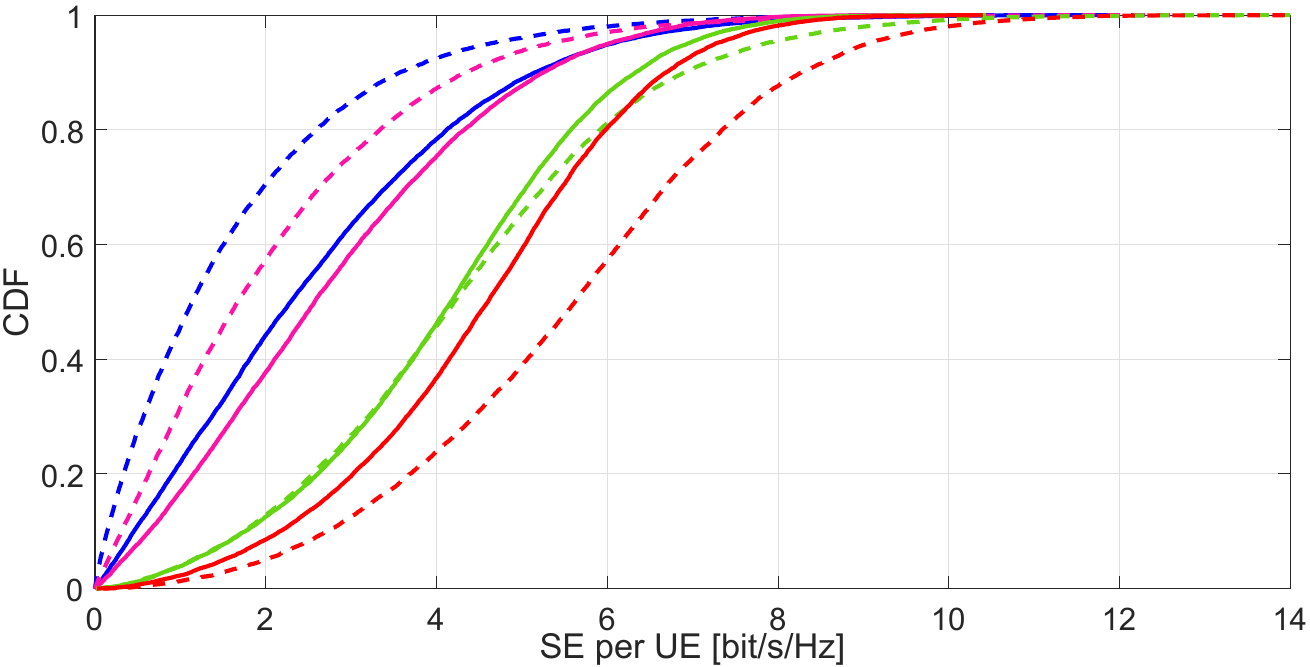}}
	\caption{Spectral efficiency of CF-mMIMO with different AP selection strategies for different city areas: raytracing vs. log-distance channel model ($Q=5$).}
	\label{Q5}
\end{figure*}

\begin{figure}[!tb] 
	\centering
	\includegraphics[width=1\columnwidth]{legend1.pdf}
	\subfigure[Amsterdam]{
		\label{apQ5.sub.1}
		\includegraphics[width=0.48\linewidth]{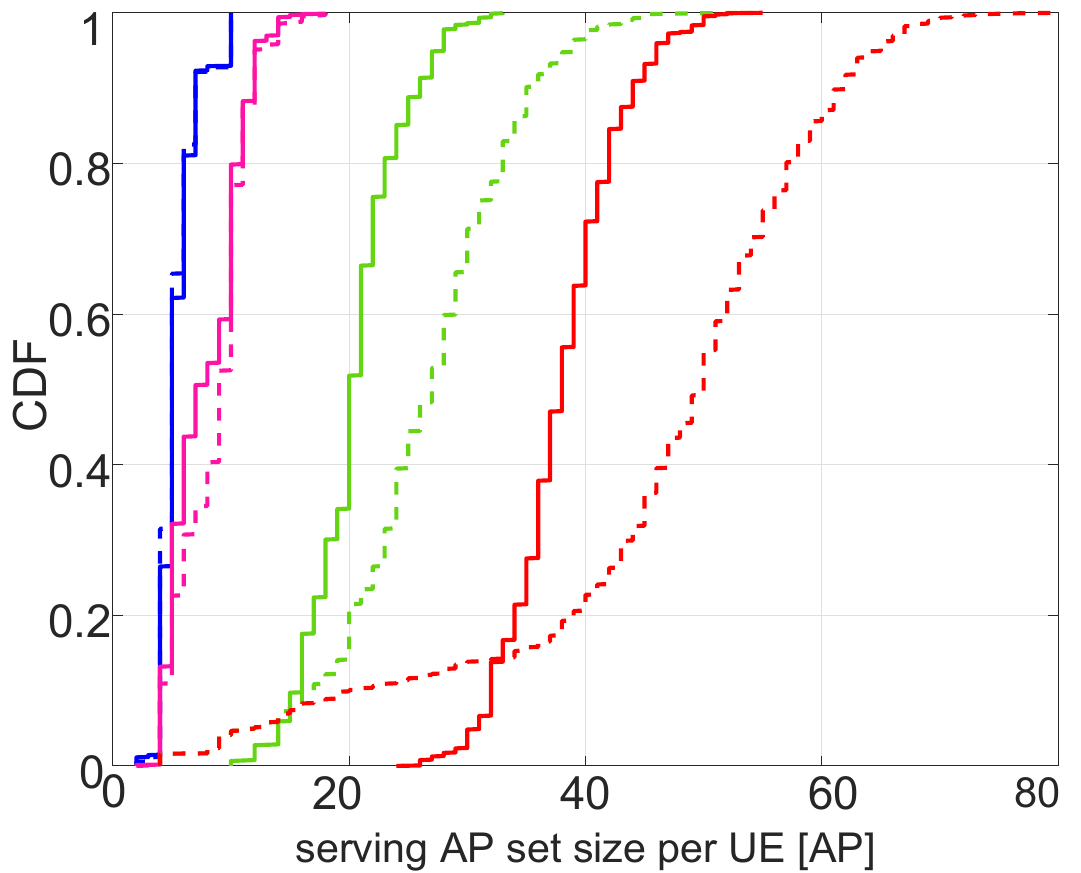}}
	\subfigure[Dresden]{
		\label{apQ5.sub.2}
		\includegraphics[width=0.48\linewidth]{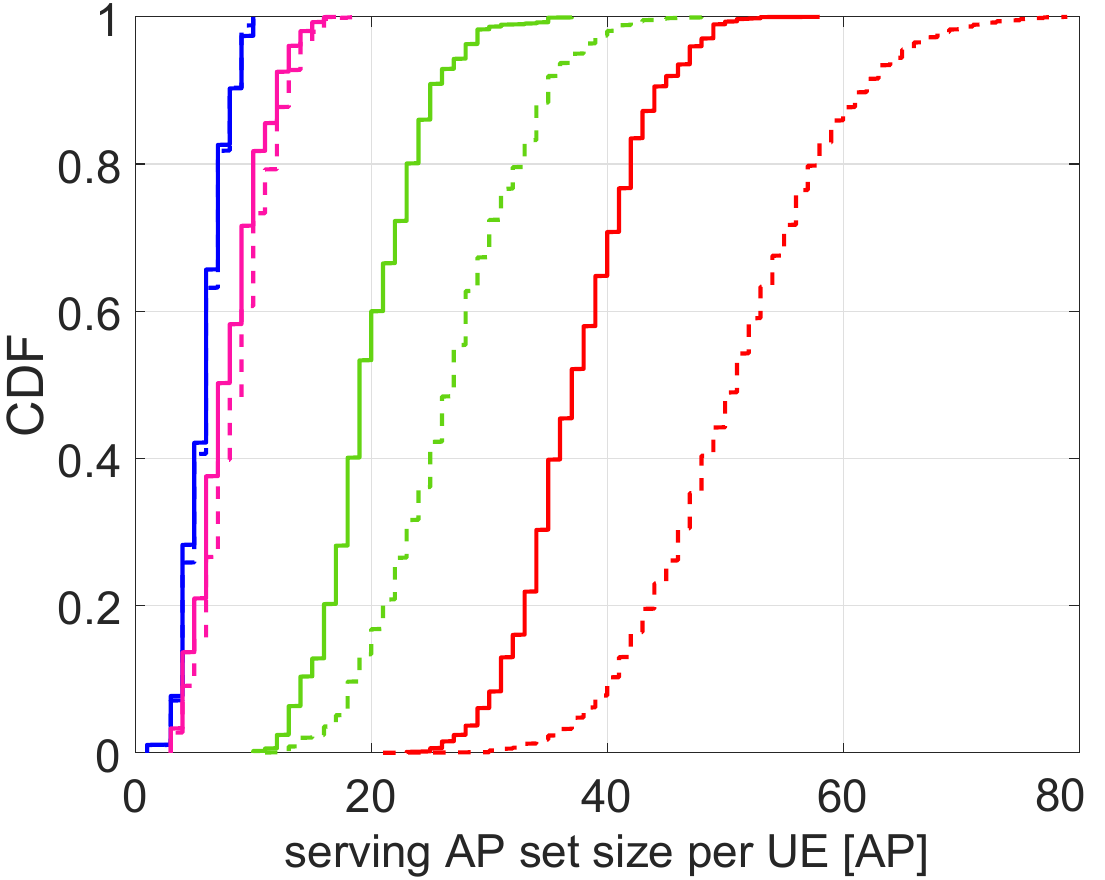}}
	\caption{Serving set size of CF-mMIMO with different AP selection strategies for different city areas: raytracing vs. log-distance channel model ($Q=5$).}
	\label{apQ5}
\end{figure}

\begin{figure*}[!tb]
\vspace{0.05cm} 
	\centering
	\subfigure[Amsterdam, raytracing channel, well-served ($5^\text{th}$\%-ile) UE]{
		\label{serveSet.AmsGood.real}
		\includegraphics[width=0.23\linewidth]{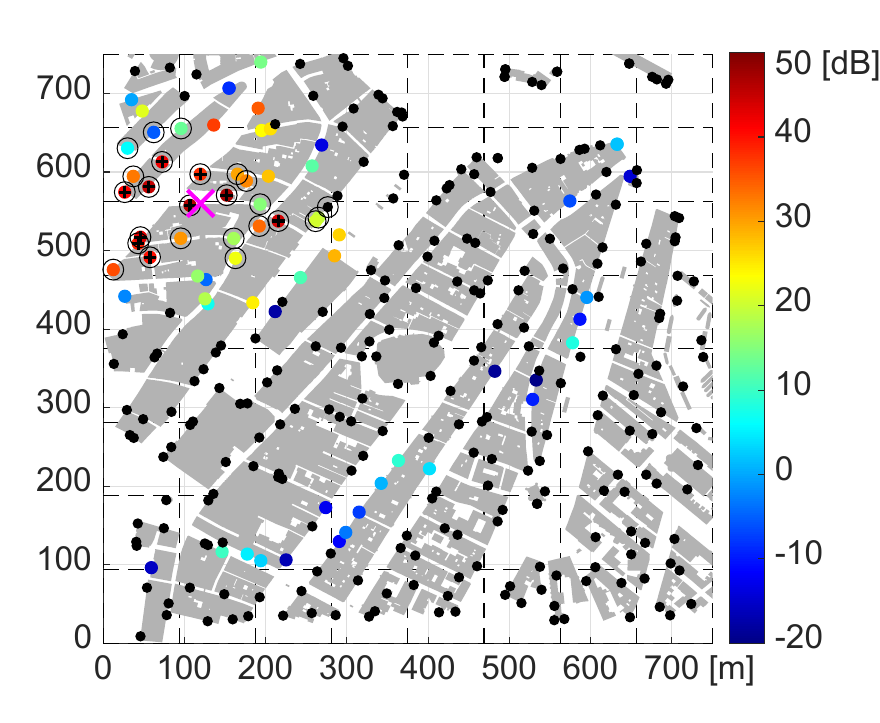}}
	\subfigure[Amsterdam, log-distance channel, same UE as Fig. \ref{serveSet.AmsGood.real}]{
		\label{serveSet.AmsGood.log}
		\includegraphics[width=0.23\linewidth]{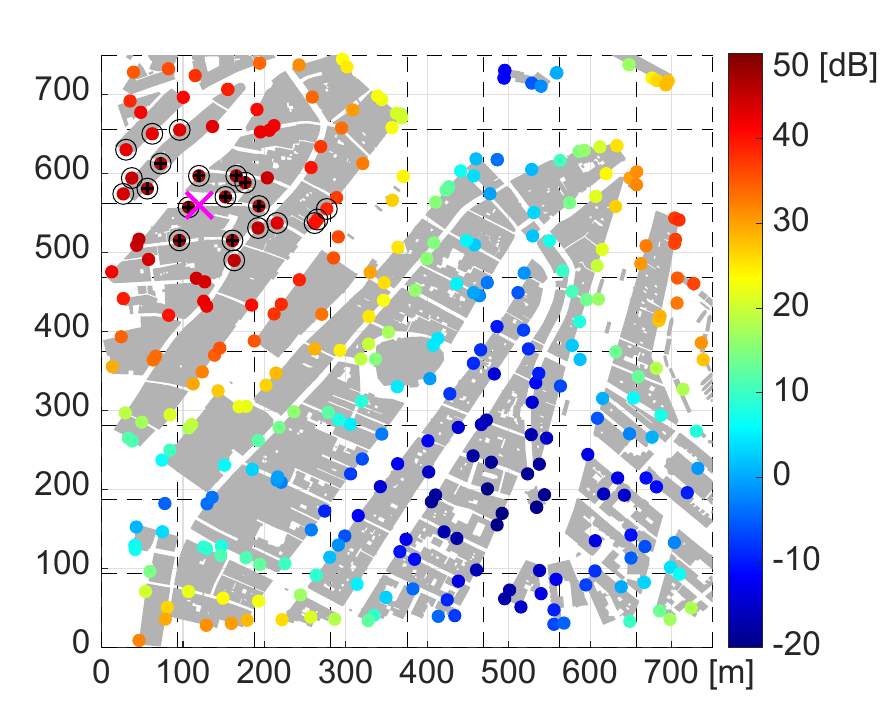}}
	\subfigure[Amsterdam, raytracing channel, worst-served \mbox{($95^\text{th}$\%-ile) UE}]{
		\label{serveSet.AmsBad.real}
		\includegraphics[width=0.23\linewidth]{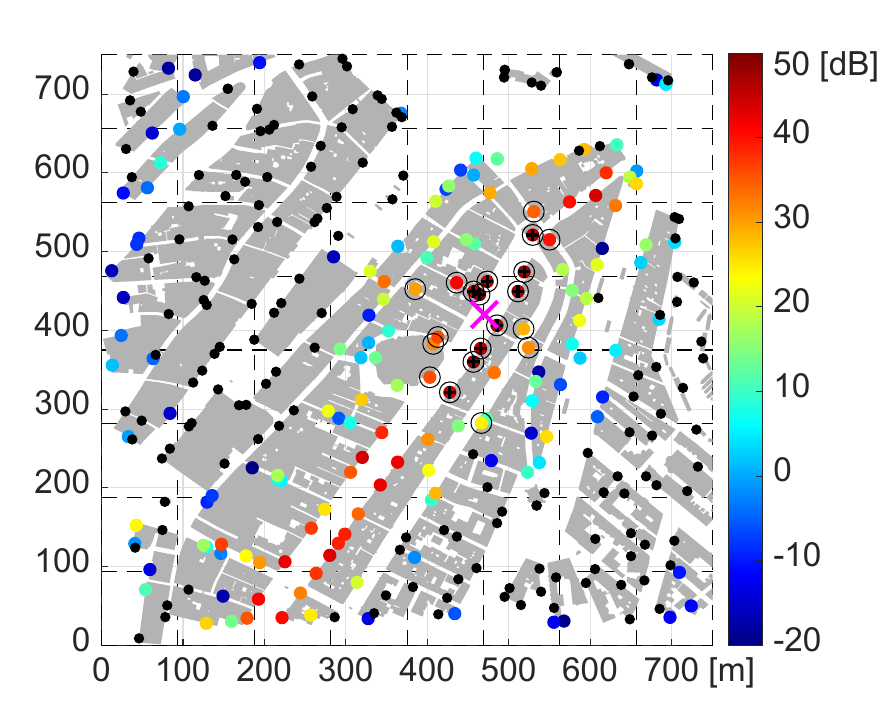}}
	\subfigure[Amsterdam, log-distance channel, same UE as Fig. \ref{serveSet.AmsBad.real}]{
		\label{serveSet.AmsBad.log}
		\includegraphics[width=0.23\linewidth]{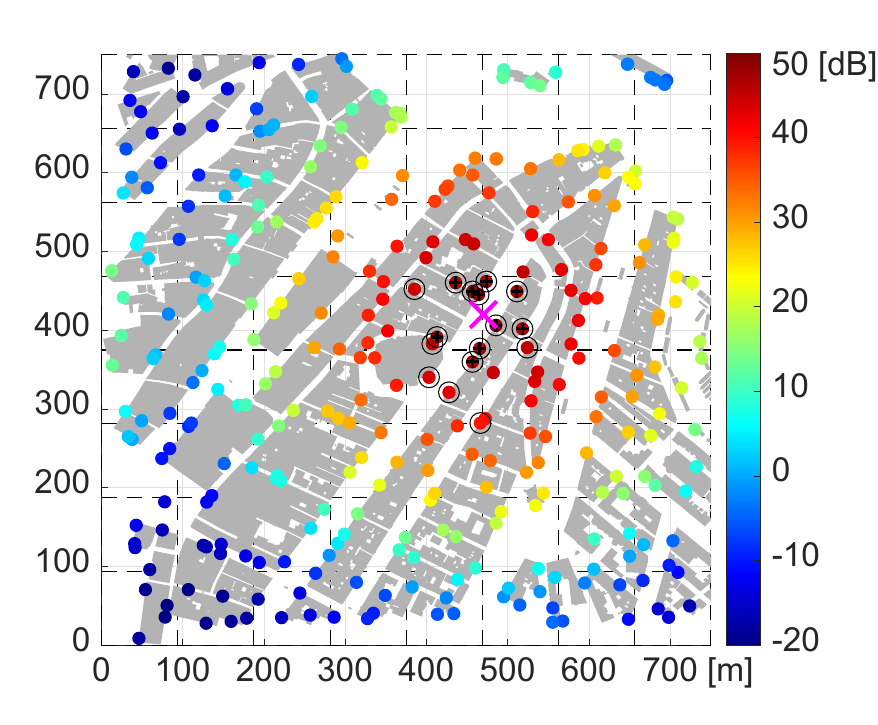}}
	\subfigure[Dresden, raytracing channel, \mbox{well-served} ($5^\text{th}$\%-ile) UE]{
		\label{serveSet.DreGood.real}
		\includegraphics[width=0.23\linewidth]{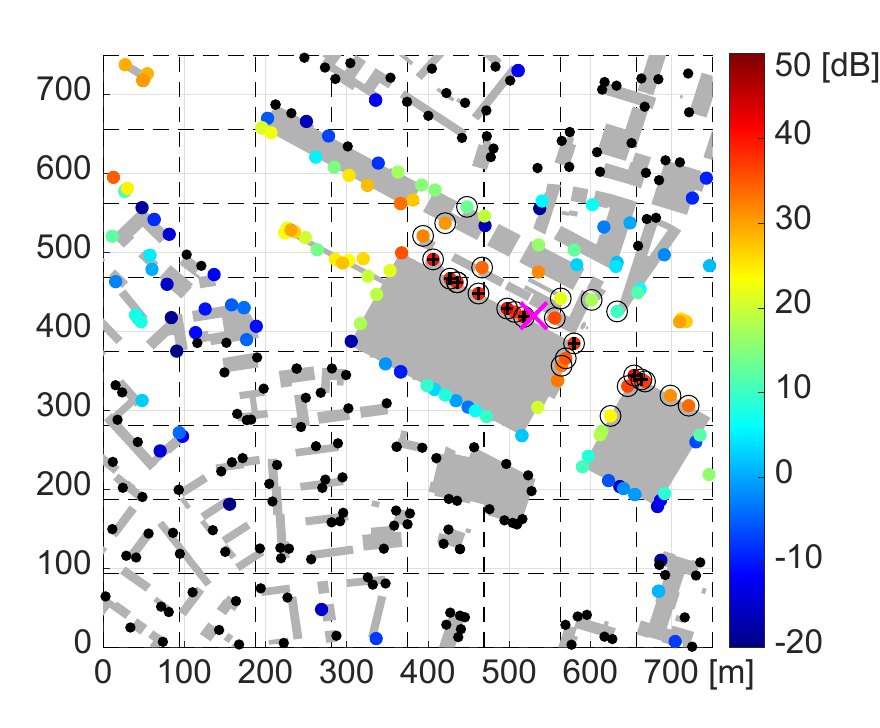}}
	\subfigure[Dresden, log-distance channel, same UE as Fig. \ref{serveSet.DreGood.real}]{
		\label{serveSet.DreGood.log}
		\includegraphics[width=0.23\linewidth]{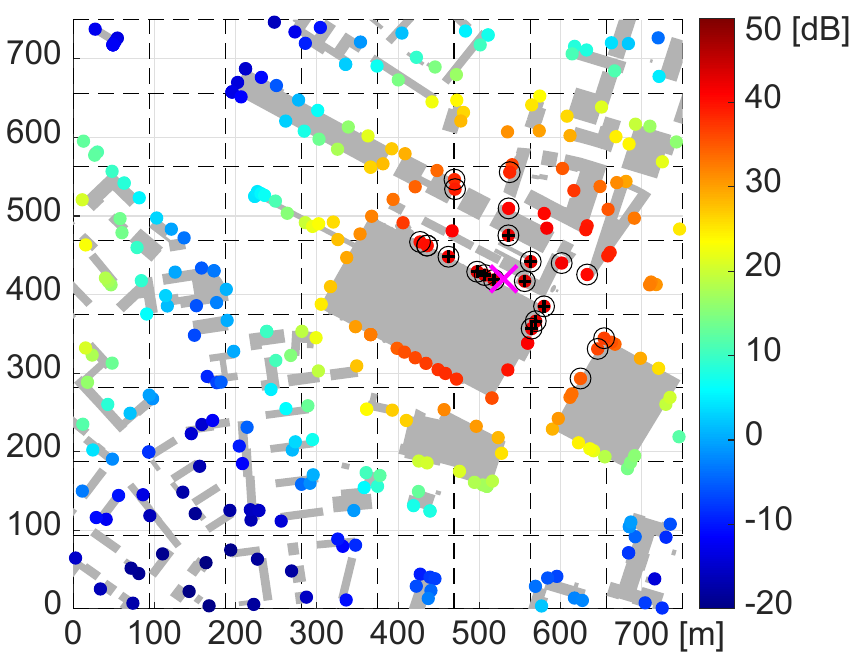}}
	\subfigure[Dresden, raytracing channel, worst-served \mbox{($95^\text{th}$\%-ile) UE}]{
		\label{serveSet.DreBad.real}
		\includegraphics[width=0.23\linewidth]{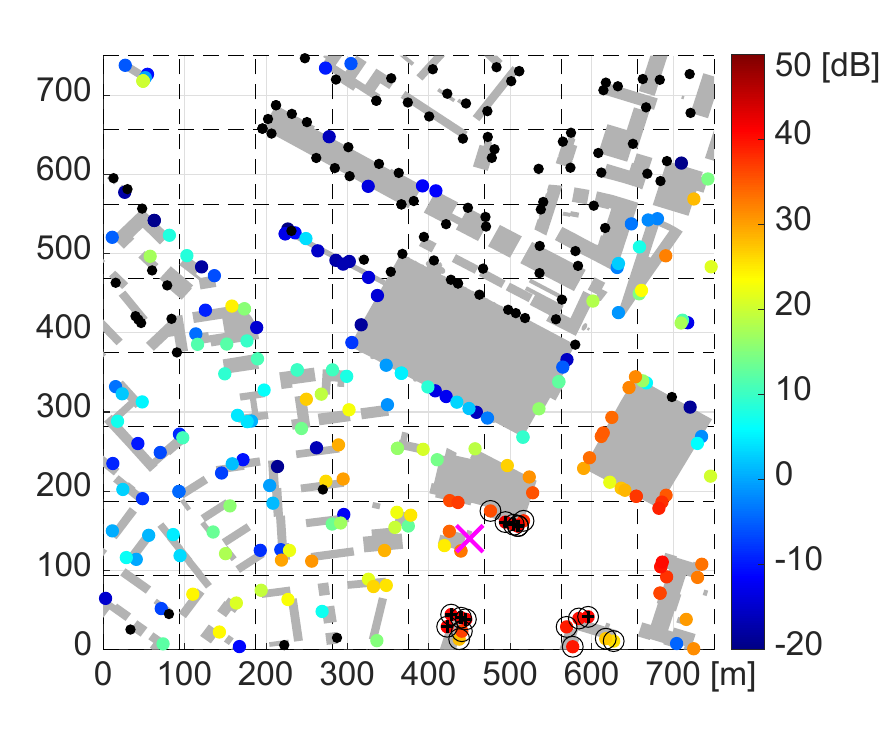}}
	\subfigure[Dresden, log-distance channel, same UE as Fig. \ref{serveSet.DreBad.real}]{
		\label{serveSet.DreBad.log}
		\includegraphics[width=0.23\linewidth]{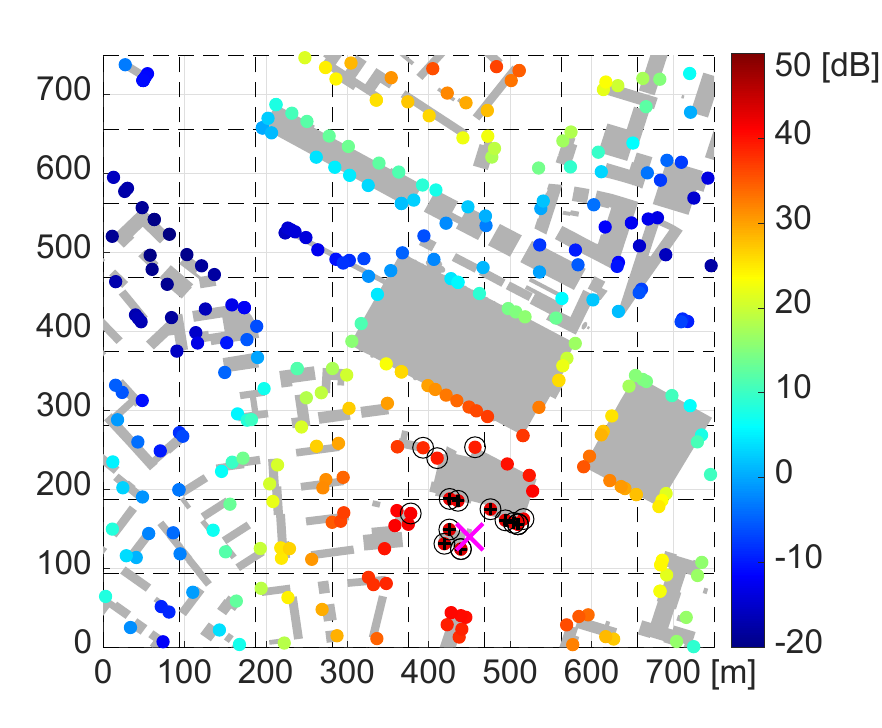}}
	\caption{SNR of each AP and the serving AP set for example UEs (pink crosses) in different cities with different channel models, with $Q=5$ and $E=10$ (APs marked by filled circles and the SNR values shown by colors, the APs with SNR lower than -20 dB marked by black dots, the $E$ APs with the best channels marked by black crosses, the serving AP set marked by black circles, and the CPU cluster boundary marked by dashed lines).}
	\label{serveSet}
\end{figure*}

\begin{figure}[!tb] 
	\centering
	\subfigure[Amsterdam]{
		\label{snrcdf.sub.1}
		\includegraphics[width=0.48\linewidth]{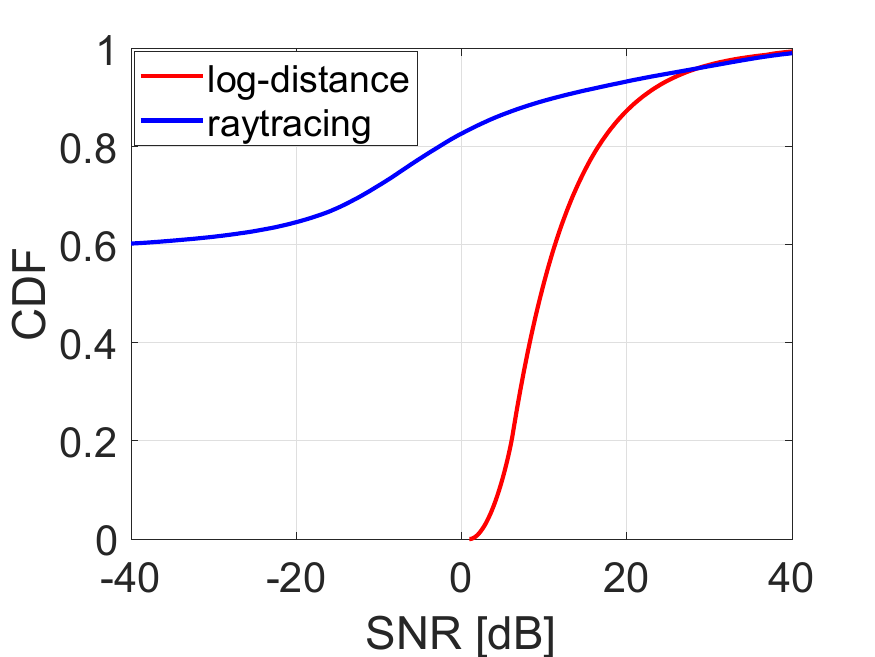}}
	\subfigure[Dresden]{
		\label{snrcdf.sub.2}
		\includegraphics[width=0.48\linewidth]{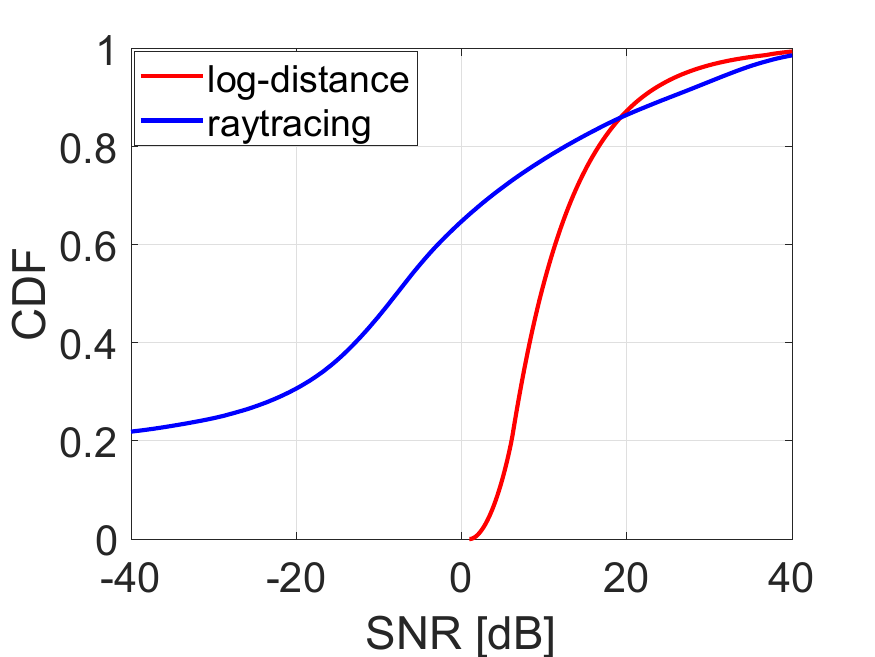}}
	\caption{SNR of AP-UE links, with different cities and channel models.}
	\label{snrcdf}
\end{figure}
\begin{figure*}[!tb] 
\vspace{0.05cm}
	\centering
	\subfigure[Amsterdam, median SE]{
		\label{rateOther.AmR}
		\includegraphics[width=0.23\linewidth]{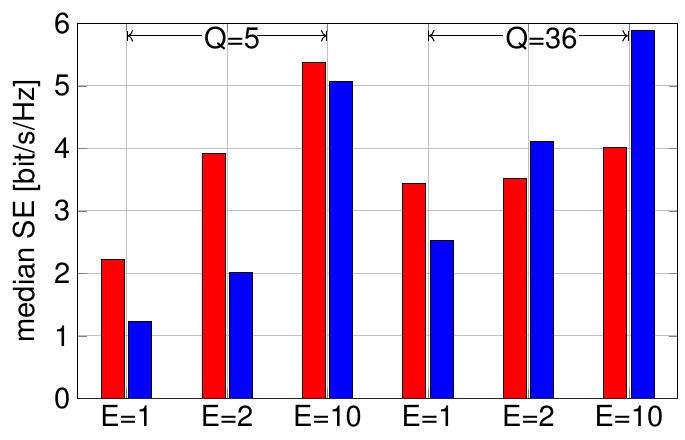}}
	\subfigure[Amsterdam, 95\%-likely SE]{
		\label{rateOther.AwR}
		\includegraphics[width=0.23\linewidth]{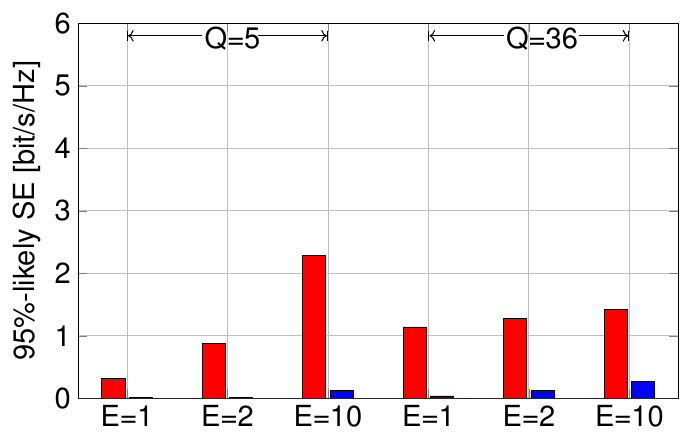}}
	\subfigure[Dresden, median SE]{
		\label{rateOther.DmR}
		\includegraphics[width=0.23\linewidth]{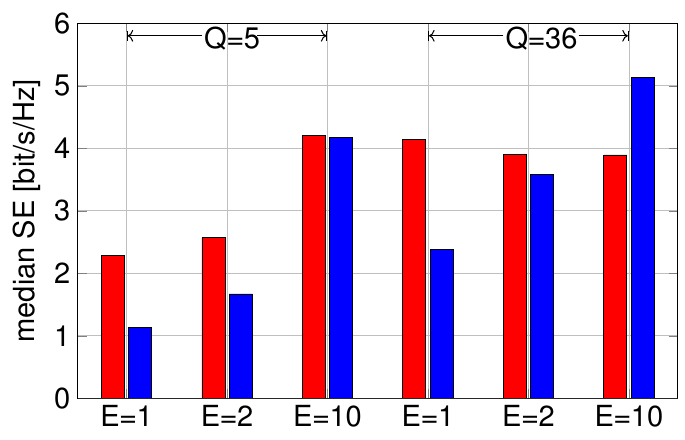}}
	\subfigure[Dresden, 95\%-likely SE]{
		\label{rateOther.DwR}
		\includegraphics[width=0.23\linewidth]{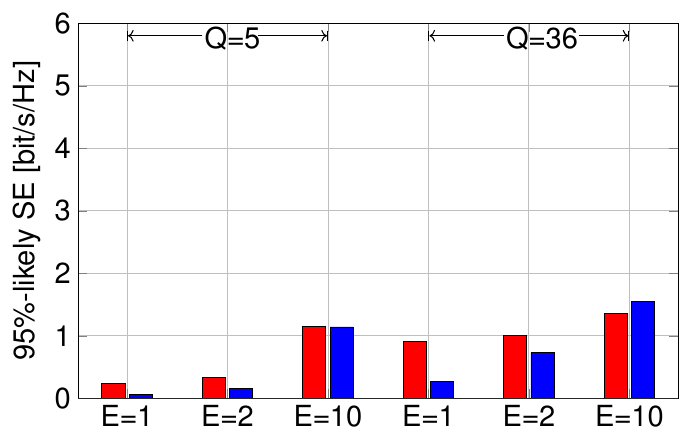}}
	\subfigure[Amsterdam, serving set size of the median-performing UEs]{
		\label{rateOther.AmD}
		\includegraphics[width=0.23\linewidth]{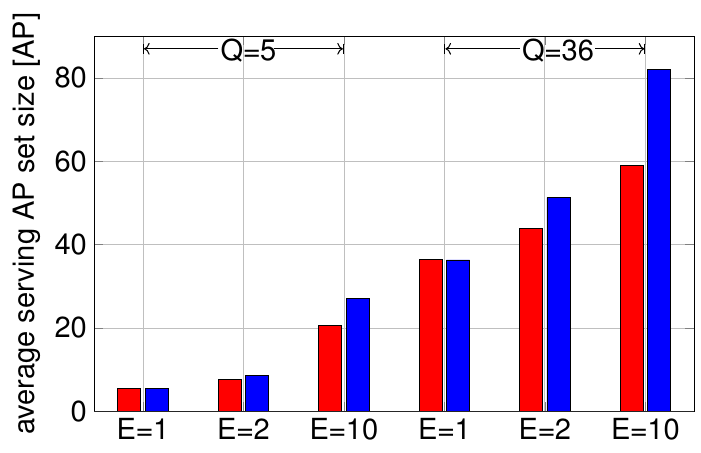}}
	\subfigure[Amsterdam, serving set size of the 95\%-likely performing UEs]{       \label{rateOther.AwD}
		\includegraphics[width=0.23\linewidth]{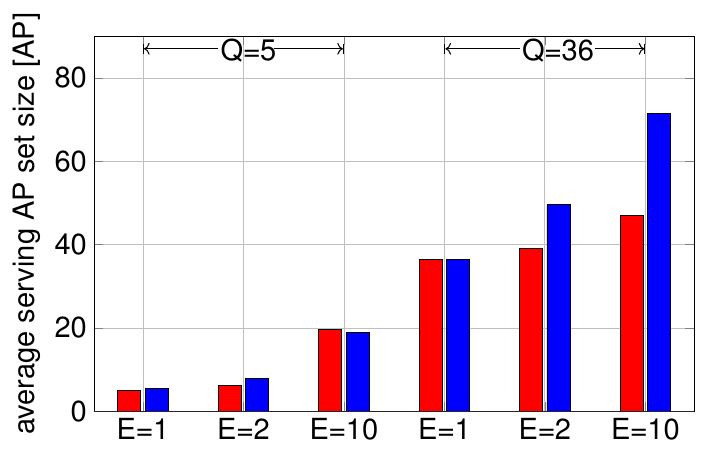}}
	\subfigure[Dresden, serving set size of the median-performing UEs]{
		\label{rateOther.DmD}
		\includegraphics[width=0.23\linewidth]{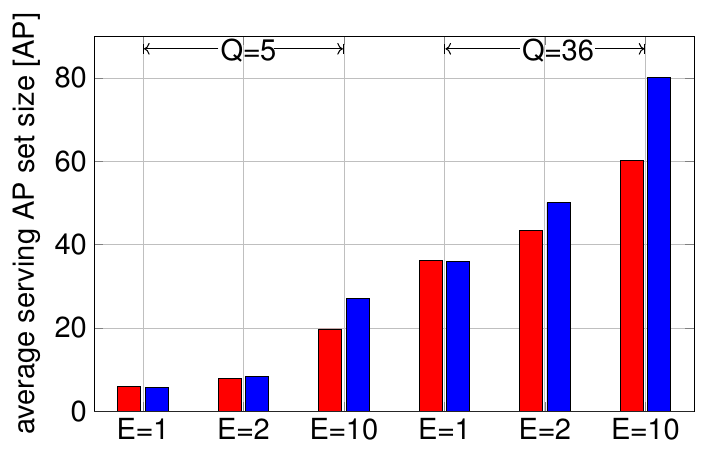}}
    \subfigure[Dresden, serving set size of the 95\%-likely performing UEs]{
		\label{rateOther.DwD}
		\includegraphics[width=0.23\linewidth]{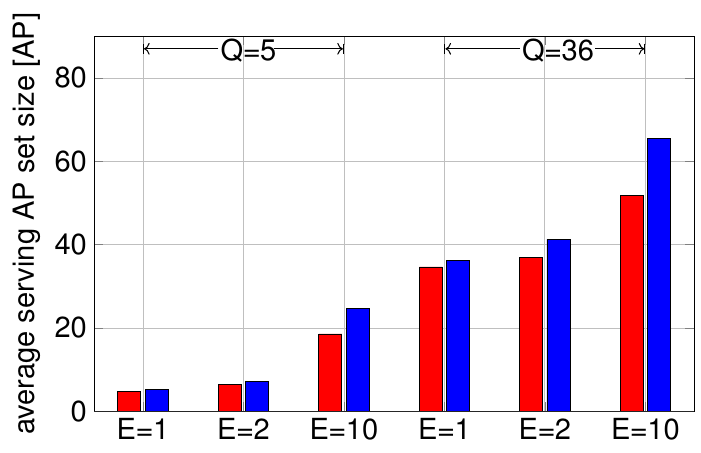}}
	\caption{SE and average serving set size of different city areas, channel models, and AP selections (red bars for log-distance and blue bars for raytracing).}
	\label{rateOther}
\end{figure*}
In this section we conduct a comparative study of the throughput performance of UE-centric CF-mMIMO in realistic urban networks under the raytracing and log-distance channel models (\textit{c.f.} Sec. \ref{channel}). We consider a network area \mbox{$S=750$ m $ \times 750$ m} in Amsterdam and Dresden, with the system model in Sec. \ref{model}. We divide the area $S$ into $8 \times 8$ disjoint square CPU clusters, corresponding to CPU cluster size of $Q=5$. We consider different $E$ values to obtain different network architectures. We obtain the SE as a distribution over all UEs with Monte Carlo simulations of 1000 realizations of UE locations. To avoid border effects, we only consider the throughput performance of the UEs that within the area \mbox{$S'=550$ m $ \times 550$ m} that has the same centroid as the area $S$. \looseness=-1

Fig. \ref{Q5} shows the SE and Fig. \ref{apQ5} shows the serving AP set size distribution of the two city areas with different channel models and serving AP selection strategies (i.e., different $E$). We also show the spatial distribution of SNR for an example UE in Fig. \ref{serveSet} and the SNR distribution of all AP-UE links in Fig. \ref{snrcdf}, to help explain the performance in detail. Let us first consider the SE performance in the city of Amsterdam in Fig. \ref{seQ5.sub.1}. With the increase of $E$, the SE and the serving set size (\mbox{Fig. \ref{apQ5.sub.1}}) with both channel models increases as more CPU clusters are included in the serving set, resulting in a higher desired signal from the serving APs and lower interference from the \mbox{non-serving} APs. However, with the same AP selection strategy, i.e., the same $E$, the log-distance and raytracing channel model obtain different SE performance. With a small $E$ (1 or 2), the majority of UEs (around 90\%) with the raytracing channel model achieve lower throughput performance compared to the log-distance model, despite the similar serving set size shown in Fig. \ref{apQ5.sub.1}. This is because in this case, the network architecture is equivalent or close to network-centric MIMO, where the UE is served by the CPU cluster that it resides in. Therefore, the serving set of a given UE is likely to be the same CPU cluster with both channel models. However, Fig. \ref{serveSet} shows that even for the same \mbox{AP-UE} link, the SNR is different due to the different channel model. Take the UE and the CPU cluster it resides (center at $[x,y]=[150,500]$) in Figs. \ref{serveSet.AmsGood.real} and \ref{serveSet.AmsGood.log} as an example: \mbox{Fig. \ref{serveSet.AmsGood.log}} shows that with the log-distance model, the SNR of all APs in this cluster is higher than 40 dB due to short AP-UE distances. By contrast, Fig. \ref{serveSet.AmsGood.real} shows that with the raytracing model, the same APs obtain significantly lower SNR than with the \mbox{log-distance} model, due to building shadowing, resulting in the lower throughput performance in Fig. \ref{seQ5.sub.1}. With a large $E$ (10 or 23), Fig. \ref{seQ5.sub.1} shows that the poorly-performing UEs with the raytracing model achieve lower SE compared to the log-distance model while the \mbox{well-performing} UEs exhibit the opposite trend. To explain the reason for this, we show the serving set for a well-performing UE (ca. $5^\text{th}$\%-ile) in \mbox{Figs. \ref{serveSet.AmsGood.real}} and \ref{serveSet.AmsGood.log} and that of a poorly-performing UE (ca. $95^\text{th}$\%-ile) in \mbox{Figs. \ref{serveSet.AmsBad.real}} and \ref{serveSet.AmsBad.log}. For the well-performing UE, the serving AP set size with the raytracing channel model is larger than that with the log-distance model, as Fig. \ref{apQ5.sub.1} shows. This is because the spatial SNR distribution is highly non-uniform with the raytracing model, which means that the $E$ APs with the best channel conditions are likely to be scattered far apart in the network and thus a larger number of CPU clusters are included in the serving set. By contrast, with the log-distance model, the APs with the best channel conditions are likely to be close to the UE and even within the same cluster. Therefore, the serving set includes fewer CPU clusters than with the raytracing model. For this reason, the well-performing UEs with the raytracing channel model obtain higher desired signal strength despite some serving APs providing lower SNR than with the log-distance model. Furthermore, the interference from the non-serving APs is lower with the raytracing model than with the log-distance model, because most of the non-serving APs provide very low SNR due to building shadowing with the raytracing model. For a \mbox{poorly-performing} UE, comparing Fig. \ref{serveSet.AmsBad.log} to \ref{serveSet.AmsBad.real} shows that even with a similar serving set size, the serving APs with the raytracing model provide lower SNR and the non-serving APs on the same street provide higher interference than with the log-distance model. Different from the log-distance channel model, where the APs within the same CPU cluster often provide similar SNR, the SNR of APs in the same CPU cluster significantly varies with the raytracing model due to building shadowing. As a result, even when the $E$ APs with the best channels may provide good SNR to the served UE, the other APs that belong in the chosen CPU cluster may not. Therefore, the poorly-performing UEs with the raytracing model obtain lower SE than with the log-distance model despite the similar serving set size. Furthermore, \mbox{Fig. \ref{seQ5.sub.1}} shows that even with the largest considered $E$, nearly half of the UEs obtain lower throughput than with log-distance model, despite more than 80\% of the UEs getting a higher serving set size (\textit{c.f.} Fig. \ref{apQ5.sub.1}). This is because as Fig. \ref{snrcdf.sub.1} shows, more than 90\% of the AP-UE links with the raytracing channel model obtain a lower SNR than with log-distance model and nearly 60\% of the links obtain an SNR lower than -40 dB, which cannot be decoded, due to building shadowing. 

Let us now study the performance in the city of Dresden. Fig. \ref{seQ5.sub.2} shows the SE in Dresden with the same serving AP selection strategies as Amsterdam in Fig. \ref{seQ5.sub.1}. Compared to Amsterdam, the SE for the same $E$ value with different channel models is more similar for Dresden. Especially when $E=10$, the throughput of the two models is nearly the same. This is because in contrast to Amsterdam, where there are many large building blocks and narrow streets, the study area in Dresden has many open areas and small building blocks, resulting in fewer diffraction and reflection effects. Therefore, the raytracing channel model in Dresden is more similar to the log-distance model. As a result, for both the example well-performing UE and the poorly-performing UE shown in Figs. \ref{serveSet.DreGood.real} - \ref{serveSet.DreBad.log}, the serving set size and SNR from the serving and non-serving APs are similar with the log-distance and raytracing models, despite the different serving APs. Nonetheless, Fig. \ref{snrcdf.sub.2} shows that with the raytracing model, there are still 80\% of the \mbox{AP-UE} links providing lower SNR than with the log-distance model due to the building shadowing. Therefore in Dresden, with a small serving set size (i.e., $E=1$ or 2) the SE with the raytracing model is consistently lower than that with the log-distance model. Only with a very large $E=23$ is the throughput performance of all UEs with the raytracing channel model higher than with the log-distance model, due to the much higher serving set size shown in Fig. \ref{apQ5.sub.2}. This illustrates that, even in an urban area with a more uniform propagation environment better represented by a log-distance path loss model, CF-mMIMO can only achieve high and uniform throughput performance with a very large serving set. However, a very large serving set leads to large signalling overhead and energy and economic costs due to the cooperation of a large number of CPU clusters, which reduces the throughput efficiency \cite{myPIMRC} and is thus likely not attractive in practice.


Let us now consider a cluster division of $3 \times 3$, resulting in $Q=36$ to represent large CPU clusters and conduct a comparative study to the previously assumed $Q=5$, which is a rather small cluster size. Fig. \ref{rateOther} shows the SE and average serving set size for the 95\%-likely and median-performing UEs in all considered scenarios. Figs. \ref{rateOther.AmD} - \ref{rateOther.DwD} show that for $Q=36$, the serving set size also increases with $E$, as for $Q=5$, and thus the SE for both median and \mbox{worst-performing} UEs in both cities increase with $E$ (\textit{c.f.} Figs. \ref{rateOther.AmR} - \ref{rateOther.DwR}). However, with $Q=36$, the SE increase with the increase of $E$ is not as significant as with $Q=5$. This is because a large $E$ and $Q$ lead to a very large overall serving set that includes \mbox{far-away} APs with a low SNR, which cannot provide high desired signal strengths. Furthermore, Figs. \ref{rateOther.AmD} - \ref{rateOther.DwD} show that the average serving set sizes of the median and \mbox{poorly-performing} UEs are similar with the same $\{Q, E\}$ combination, since the serving set size is decided by the $\{Q, E\}$ combination for all UEs. This illustrates that for the worst-performing UEs, the low SE is not caused by the small number of serving APs but rather the low SNR provided by the serving set. The serving APs are either too far away from the UE (log-distance channel) or are blocked by buildings (raytracing channel).

Comparing the two channel models, Fig. \ref{rateOther.AmR} shows that for Amsterdam, the median SE with the raytracing model can only surpass that of with the log-distance model with $Q=36$ and $E=2$ or 10, i.e., a very large serving set size, due to a much higher obtained serving set size shown in \mbox{Fig. \ref{rateOther.AmD}}. For the \mbox{worst-performing} UEs, Fig. \ref{rateOther.AwR} shows that for all considered cases with the raytracing channel model, the SE is nearly zero and the UEs are in outage due to the building shadowing illustrated in Fig. \ref{serveSet.AmsBad.real}, regardless the large serving set size shown in Fig. \ref{rateOther.AwD}. This shows that in a highly \mbox{non-uniform} urban propagation environment such as Amsterdam, \mbox{UE-centric} \mbox{CF-mMIMO} cannot provide good performance for the \mbox{worst-performing} UEs as originally claimed \cite{cfvs, cfsim, scalableCF}, even with a very large serving set size. For Dresden, where the propagation environment is more uniform and thus better represented by the log-distance model, Figs. \ref{rateOther.DmR} and \ref{rateOther.DwR} show that the raytracing channel model obtains comparable and even higher SE than the log-distance model for both median and worst-performing UEs only when $E=10$, i.e., for a large serving set size (\textit{cf.} Figs. \ref{rateOther.DmD} and \ref{rateOther.DwD}), which results in too high signaling overhead and cost \cite{myPIMRC}. Fig. \ref{rateOther} thus emphasizes that when considering the raytracing channel model representative of the realistic propagation environment, CF-mMIMO cannot provide as high throughput performance as observed with the log-distance channel in the existing literature, especially for the worst-performing UEs. This suggests that site-specific and propagation environment-aware network architecture design is required for CF-mMIMO to be attractive in practice.

\section{Conclusions}
\label{conclude}
We presented the first comparative performance analysis of CF-mMIMO in realistic urban network propagation environments, under both the log-distance and raytracing channel models. We analysed the distribution of SNR and SE of \mbox{CF-mMIMO} in two representative city areas, Amsterdam and Dresden, under different network architectures. Our results show that with the raytracing channel model representative of the realistic urban propagation environment, \mbox{UE-centric} \mbox{CF-mMIMO} cannot provide as high throughput performance as observed with the log-distance channel widely assumed by the existing literature, especially for the worst-performing UEs. This puts into question whether CF-mMIMO can deliver the promise of high and uniform throughput performance in practice, unless novel site-specific and propagation \mbox{environment-aware} network architectures are designed, which is the focus of our ongoing work.

\vspace{-0.1cm}
\bibliographystyle{IEEEtran}
\bibliography{IEEEabrv,IEEEexampleMaster}

\end{document}